\documentstyle[11pt,moriond,epsfig]{article}

\def\Journal#1#2#3#4{{#1} {\bf #2}, #3 (#4)}

\def\PLB{{\em Phys. Lett.}  B}

\def\JPC{{\em Eur. Phys. J.} C}
\def\JPD{{\em Eur. Phys. J. direct} C}
\def\JPG{{\em Eur. Phys. J.} G}

\def\be{\begin{equation}}
\def\ee{\end{equation}}
\def\bea{\begin{eqnarray}}
\def\eea{\end{eqnarray}}

\begin{document}
\title{Jets and Event Shape Studies in \boldmath{$ep$}-collisions at HERA}

\author{ R. P\"OSCHL }

\address{LAL, Groupe H1,\\
91898 Orsay Cedex, France\\
On behalf of the H1 and ZEUS collaborations}

\maketitle\abstracts{The results of complementary tests of perturbative 
QCD in DIS at large centre-of-mass energies at HERA are
presented. The analysis of event shape variables allows for  
the investigation of approaches to understand aspects of non-perturbative QCD 
such as the formation of the hadronic final state. The strong coupling 
constant, $\alpha_s$, is extracted from both inclusive 
jet cross sections and dijet rates and is found to be competitive 
with the world average.}

\section{Introduction}\label{sec:intro}
\begin{figure}[h]
\begin{center}
\psfig{figure=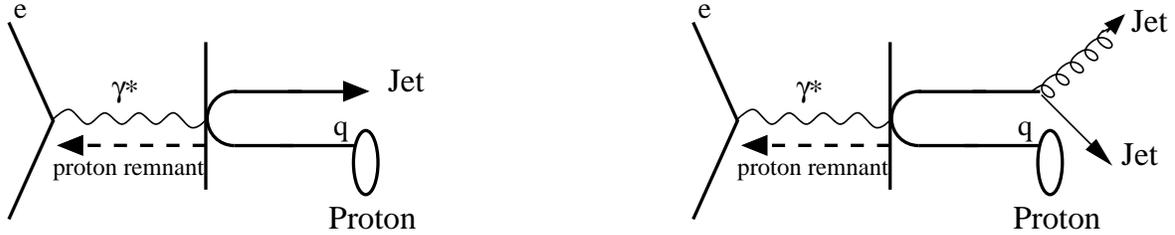,height=1.2in}
\end{center}
\caption{Deep inelastic scattering with (right) and without (left) QCD
corrections as viewed in the Breit-Frame. The QCD corrections lead
to final state jets with non-zero $E_t$. \label{breit}}
\end{figure}

The hadronic final state in deep inelastic $ep$ scattering (DIS), i.e.
$ep\rightarrow eX$, provides a rich testing ground for studies of QCD.
The HERA $ep$ collider offers the unique possibility to investigate QCD over several
orders of magnitude in the virtuality, $Q^2$, of the exchanged boson.  
Figure~\ref{breit} displays the DIS process as viewed in the
Breit-Frame of reference. 
In zeroth order the scattered quark leaves the interaction
collinear with the direction of the incoming photon. The radiation
of a gluon (Fig.~\ref{breit}), or the process $\gamma^{\star}g\rightarrow
q\bar{q}$, however, generates a non-zero transverse momentum of the final state gluon or quarks, 
resulting in distinct properties of the hadronic final state as exploited in high 
statistics analyses from the H1 and ZEUS collaborations presented in this article.

\section{Event Shapes in DIS}\label{sec:prod}

In event shape studies, the complete set of hadronic final state particles
is used to calculate proper observables which are sensitive to QCD 
corrections to the inclusive process. The analyses in the Breit-Frame of the
thrust $\tau_{m}(\tau_{c})$, measured with respect to the thrust axis, the $C$-Parameter 
and the jet mass $\rho$ ($\rho_{0}$) permits the
direct comparison of event shapes in DIS to event shapes in time-like 
$e^{+}e^{-}$-events. 
The thrust $\tau(\tau_{z})$ and the jet broadening $B(B_{c})$ given by 

\begin{minipage}{7.5cm}
\begin{eqnarray}
\tau(\tau_{z}) = 1- \frac{\sum p_{z,h}}{\sum |p_{h}|}
\label{tauz}
\end{eqnarray}
\end{minipage}
\begin{minipage}{7.5cm}
\begin{eqnarray}
B(B_{c})=\frac{\sum p_{t,h}}{\sum |p_{h}|}
\label{broad}
\end{eqnarray}
\end{minipage}
\vspace{0.2cm}

\noindent are measured with respect to the boson axis and are therefore specific to DIS.
The sum runs over the longitudinal and transverse momenta of the
hadronic final state particles projected onto the axis of the incoming
boson. From Eqs.~\ref{tauz} and  \ref{broad} and Fig.~\ref{breit} it is easy 
to see that the defined variables are zero in case of lowest order DIS and different 
from zero in the higher order case including the hadronization step.
\vspace{0.2cm}

\unitlength1.0cm
\begin{figure}[h]
\begin{minipage}[b]{5.0cm}
\begin{picture}(6.0,5.0)
\epsfig{figure=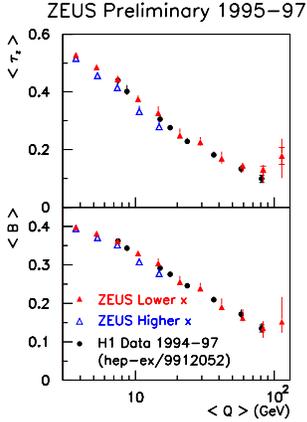,height=6.2cm}
\end{picture}\par
\caption{Means of different event shape variables as a function
of $\langle Q \rangle$ as measured by the H1 and ZEUS collaborations. \label{meanevs}}
\end{minipage}
\hfill
\begin{minipage}[b]{10.0cm}
The measured distributions of the mean 
values of $\langle \tau_{z} \rangle$ and $\langle B  \rangle$ 
are shown in Fig.~\ref{meanevs}~\cite{zsos,h1evs}. The slope of 
the distributions with $\langle Q \rangle$ indicates that the events get
more collimated with increasing boson virtuality. Here the results of the H1 and ZEUS  
collaborations agree within errors which is also for the variables not shown in
Fig.~\ref{meanevs}.
The ZEUS collaboration investigates the $x$-dependence of the event shape variables, 
where $x$ is known as the fractional momentum of the 
interacting parton with respect to the incident proton momentum. Perturbative QCD 
up to $\cal{O}$$(\alpha^{2}_{s})$ is not sufficient to describe the measured 
means of the event shape variables~\cite{zsos,h1evs} and non-perturbative effects 
such as hadronization have to be considered. The mean of any event shape 
variable, $\langle \cal{F} \rangle$, may thus be decomposed into a 
perturbatively calculable part and a part which contains the non-perturbative physics   
\begin{equation}
\langle \cal{F} \rangle  = \langle F \rangle^{\it pert.} + \langle
\cal{F} \rangle^{\it npert.}. 
\end{equation}

\end{minipage}
\end{figure}

\begin{figure}
\begin{center}
\begin{picture}(12.0,6.0)
\put(7,-0.5){\epsfig{figure=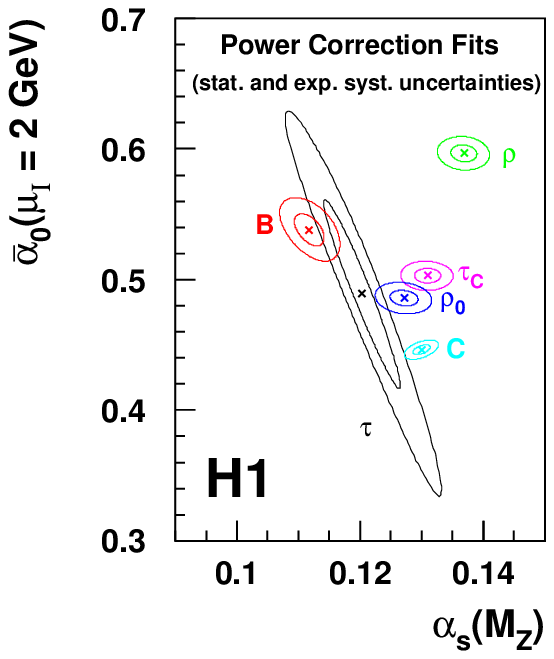}}
\put(-1,0.){\epsfig{figure=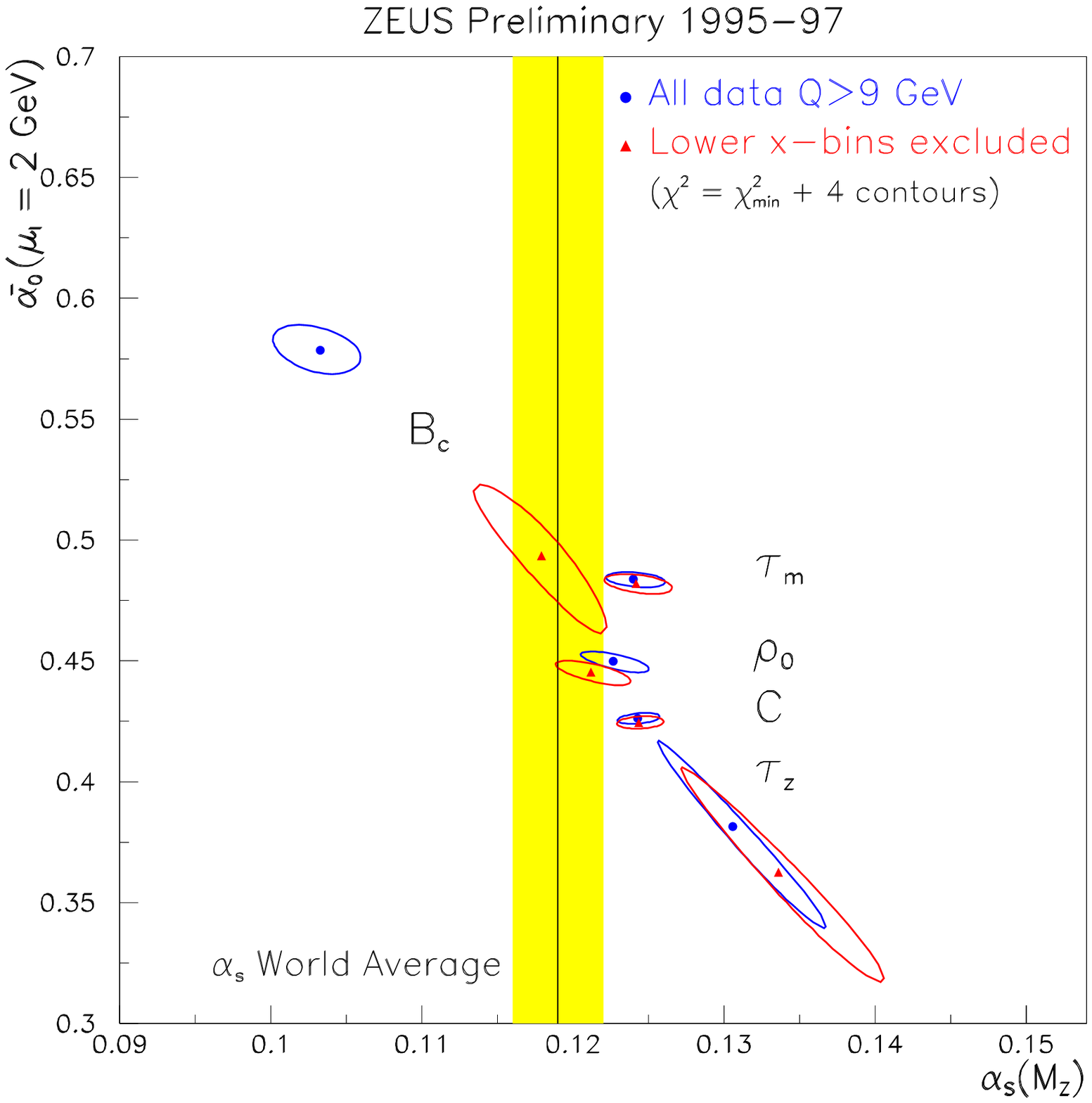,height=6.8cm}}
\put(0.5,2.5){ a) }
\put(9.5,2.5){ b) }
\end{picture}
\end{center}
\caption{Results of power correction fits to the mean values of
the event shape variables. \label{contours}} 
\end{figure}

In the analyses presented here hadronization is treated conceptually by attributing 
power corrections~\cite{dokweb} to soft gluon phenomena. In this framework, the 
non-perturbative part $\langle \cal{F} \rangle^{\it npert.}$ can be written as
\begin{equation}
\langle \cal{F} \rangle^{\it npert.} =  \it{ a_{F} \cdot {\rm 1.61} \frac{\mu_{I}}{Q}
\left[ \bar{\alpha}_{0}(\mu_{I})-\alpha_{s}(Q)-{\rm 1.22}  \left( {\rm
ln}\frac{Q}{\mu_{I}} +{\rm 1.45} \right) \alpha_{s}(Q) \right]},   
\label{dokap}
\end{equation}
with an additional enhancement $a^{\prime}_{B}$ in case of the jet broadening~\cite{dokvan}
which might also depend on $x$.
Note the free parameter $\bar{\alpha}_0$ which takes soft gluon effects
into account and can be interpreted as the mean of $\alpha_s$ in  
the region $0<\mu_{I}<2\,{\rm GeV}$. Good fits to the data of Fig.~\ref{meanevs} are obtained 
using $\bar{\alpha}_0$ and $\alpha_s$ as free parameters~\cite{h1evs,zsos}. 
The results on $\alpha_s, \bar{\alpha}_0$ 
obtained by the H1 and ZEUS collaborations are displayed in Fig.~\ref{contours} 
together with the $\chi^2_{min}+4$ (Fig.~\ref{contours} a) and $\chi^2_{min}+4(1)$ error
ellipses (Fig.~\ref{contours} b). 

\unitlength1.0cm
The $\alpha_s$ of the two analyses are broadly
consistent with each other and with the world average although the large spread
reveals the need for higher order corrections. The resulting $\bar{\alpha}_0$ 
of both analyses suggest an universal value of $0.5\pm0.1$ for the event shape 
variables $\tau_{m}(\tau_{c})$, $C$ and $\rho_{0}$.
The H1 analysis as shown in Fig.\ref{contours} b) demonstrates the influence of
the treatment of hadrons on the jet masses $\rho$ and $\rho_0$ where the assumption
of massless hadrons ($\rho_{0}$) leads to a more consistent interpretation 
of power corrections. The fit results on $B(B_{c})$ differ between the both experiments
although the measurement as shown in Fig.~\ref{meanevs} is in good agreement. 
As already indicated in Fig.~\ref{meanevs} the ZEUS collaboration splits the
data into two $x$ ranges and the resulting $\bar{\alpha_0}$ values obtained 
for the $B$-Parameter indicate a strong $x$-dependence in contrast to the weak dependence 
seen in Fig.~\ref{meanevs}. The fit results on $\tau(\tau_{z})$ exhibit large uncertainties 
and strong correlations between $\alpha_s, \bar{\alpha}_0$, making the interpretation of the
result quite difficult.

\section{\boldmath{$\alpha_s$} and Gluon Density from Jet Cross Sections }
As indicated in Fig.~\ref{breit} the radiation of gluons leads
to final state jets with non-zero transverse momentum $E_t$. In
DIS high $E_t$ jets are, in first order $\alpha_s$, produced 
by the Boson-Gluon Fusion process and the QCD-Compton process. 
The resulting jet cross section can be expressed in 
a power series of the strong coupling constant:
\begin{equation}
\sigma_{jet} = \sum \alpha^{n}_s \sum_{i=q,G} {\rm C}_{i,n} \otimes {\rm pdf}_{i}.
\label{jetcross}
\end{equation}
Here the ${\rm C}_{i,n}$ denote the matrix elements which can be
calculated in perturbative QCD and ${\rm pdf}_{i}$ denotes the parton density of
a parton of type $i=q,G$ (q=quark, G=gluon). It is
discussed in~\cite{h1as} that fundamental quantities such as $\alpha_s$ or
the pdf can be determined for $Q^2>150\,{\rm GeV}^2$ by means of a QCD analysis
due to the small uncertainties of NLO-QCD ($\cal{O}$ $(\alpha^{2}_{s})$) predictions
in this phase space region.

\begin{figure}
\begin{center}
\begin{picture}(12.0,6.)
\put(6.5,-0.5){\epsfig{figure=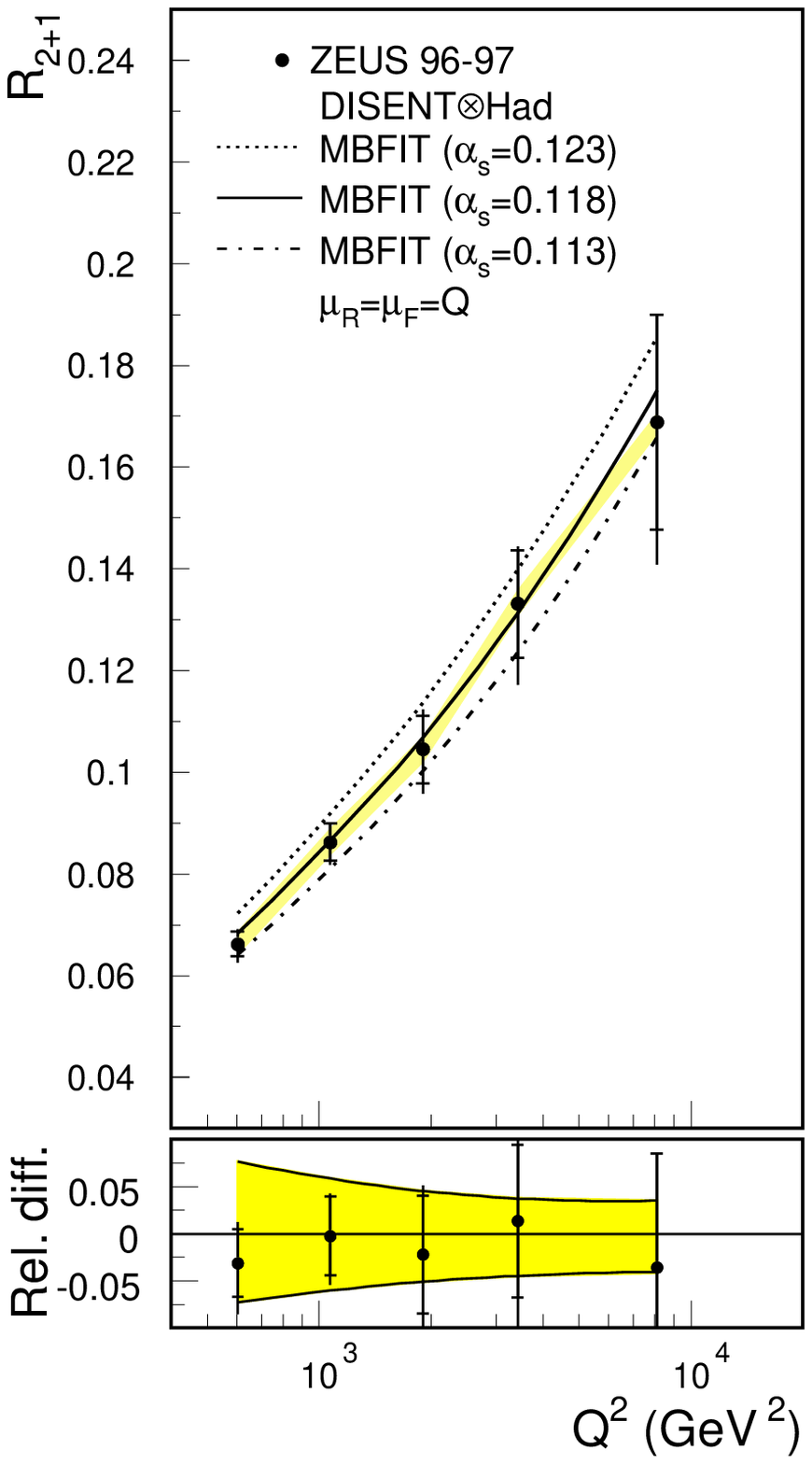,height=7.3cm}}
\put(-2,-0.5){\epsfig{figure=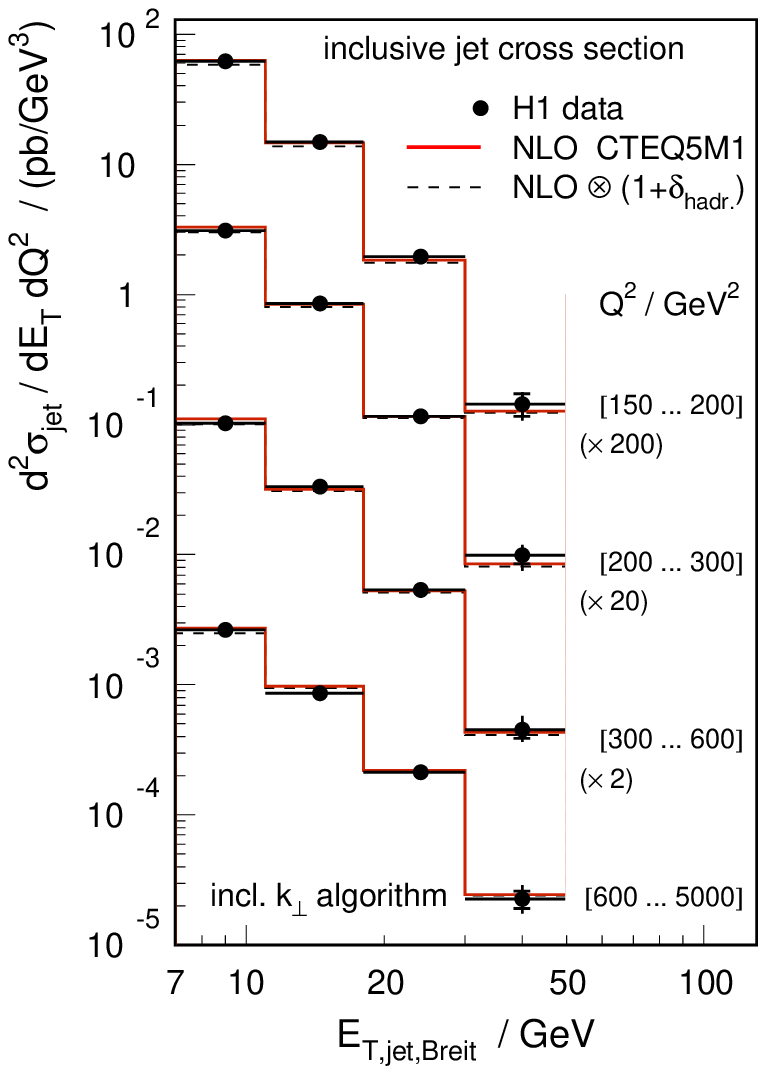,height=7.3cm}}
\end{picture}
\end{center}
\caption{Inclusive jet cross section (left) and 
dijet rates (right) used in independent determinations of $\alpha_s$ 
by the H1 and ZEUS collaborations, respectively \label{jtpl}} 
\end{figure}

Two complementary approaches have been used for the extraction of
$\alpha_s$ by H1~\cite{h1as} and ZEUS~\cite{zsas} each with slightly
different advantages. Jets were defined by using the
inclusive $k_t$-algorithm. The inclusive jet cross section as shown 
in Fig.~\ref{jtpl} (left) was 
proven to have small uncertainties from the hadronization step. The partial
cancellation of pdf uncertainties when using dijet rates, Fig.~\ref{jtpl} (right), leads 
to small uncertainties from the pdf.

\unitlength1.0cm

Both measurements are compared to NLO-QCD calculations corrected
for hadronization effects. The results on $\alpha_s$
are 
\begin{equation}
\alpha_s(M_{Z}) = 0.1186 \pm 0.0007 \; {\rm (stat.)}
\pm 0.0030 \; {\rm (exp.)} \pm 0.0051 \; {\rm (th.)}
\label{asij}
\end{equation}
from inclusive jet cross sections~\cite{h1as} and
\begin{equation}
\alpha_s(M_{Z}) = 0.1166 \pm 0.0019 \; {\rm (stat.)}
^{+ 0.0024}_{- 0.0033} \; {\rm (exp.)} ^{+ 0.0057}_{- 0.0044} \; {\rm (th.)}
\label{asr2}
\end{equation}
from dijet rates~\cite{zsas}. The results both agree with the current world
average~\cite{aswa} $\alpha_s(M_{Z})=0.1184\pm0.0031$. Note, that the experimental errors
are much smaller than the theoretical uncertainties, of which the pdf uncertainties
dominate~\cite{h1as}.

\begin{figure}
\begin{center}
\begin{picture}(12.0,6.)
\put(6.5,-0.5){\epsfig{figure=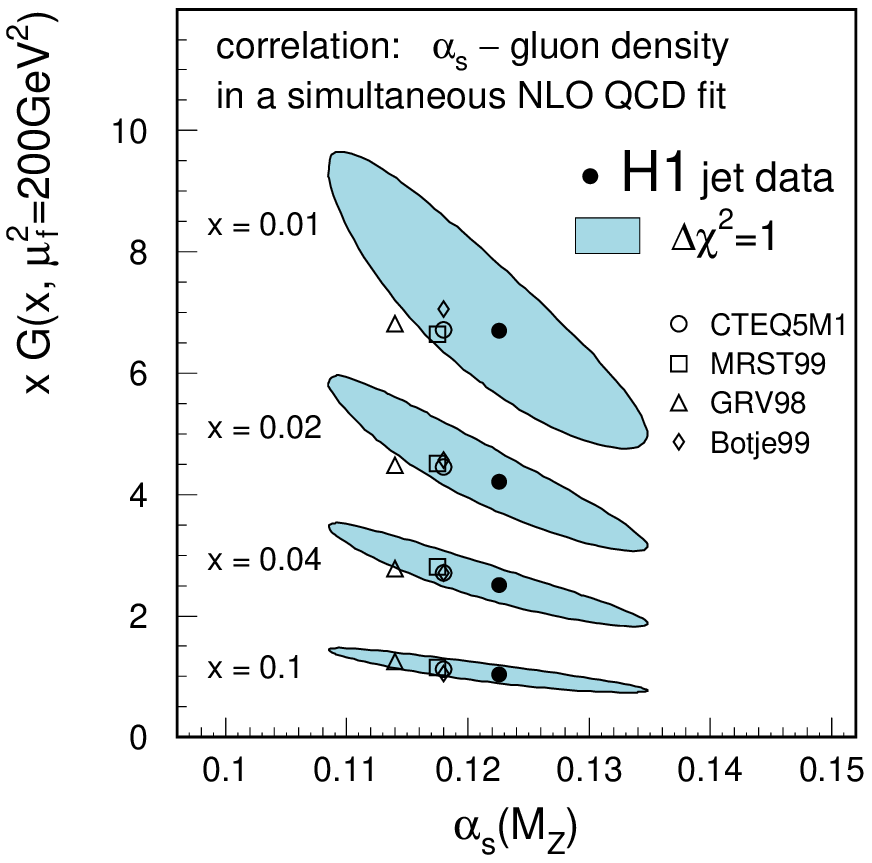,height=7.3cm}}
\put(-2,-0.5){\epsfig{figure=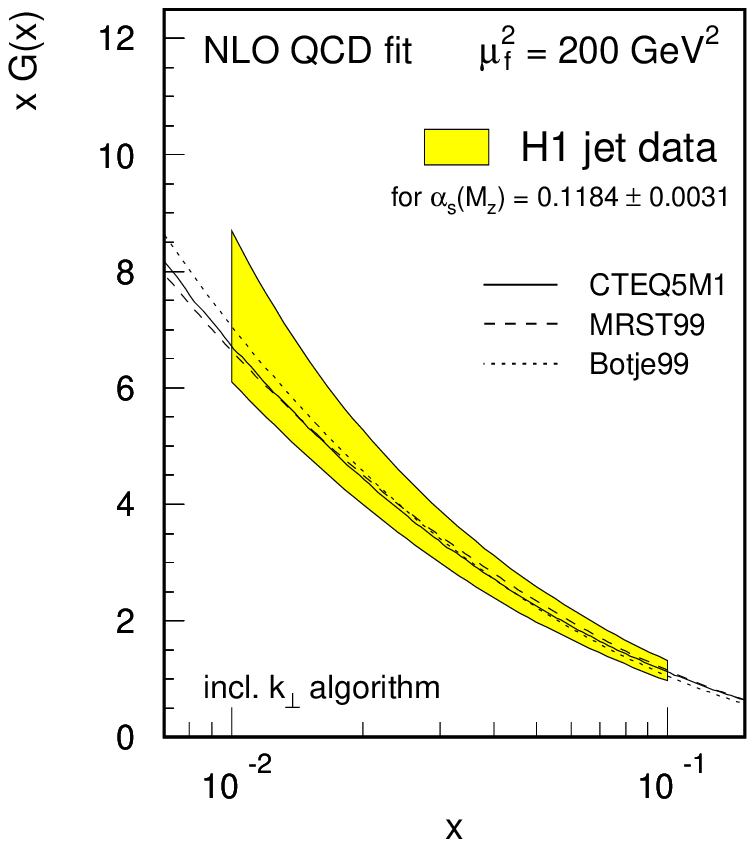,height=7.3cm}}
\end{picture}
\end{center}
\caption{Left: The gluon density $xG(x)$ as determined from DIS Data including 
jet cross sections for a fixed value of $\alpha_s$. The error band includes
the combined experimental and theoretical uncertainties.
Right: Results of a simultaneous fit of $\alpha_s$ and the gluon density in
the proton to H1 jet and inclusive DIS data. The central
fit results (black points) are displayed together with their $\chi^2_{min}+1$
error ellipses \label{asgl} } 
\end{figure}

The extraction of $\alpha_s$ needs the pdf as external input. On the
other hand, using the best knowledge of $\alpha_s$, the jet cross-sections
can be used to extract the quark and gluon densities in the proton.
While the quark densities can be well constrained by using
inclusive DIS data~\cite{f2lq}, jet cross sections are particularly sensitive to the
gluon density. Figure~\ref{asgl} (left) shows the determination of the gluon density in the 
proton where the current world average for $\alpha_s$ was used as input. 
A more independent test of QCD can be performed 
by fitting $\alpha_s$ and the quark and gluon densities simultaneously. 
Figure \ref{asgl} (right) displays the result in the form of a correlation plot 
of $\alpha_s$ versus the gluon density. 
\unitlength1.0cm
The obtained values agree with those from global fits by various other groups within
the errors given by the $\chi^2_{min}+1$ error ellipses. 
The large eccentricities of the ellipses demonstrate that the result 
is sensitive to the product $\alpha_s \cdot G(x)$
while the size of the uncertainties, however, indicates that the precision 
of the measurement is still limited.
The determination of the fundamental quantities will benefit
from NNLO-calculations and/or the analysis of three-jet events 
for which the upcoming HERA II running promises sufficient statistics.

\end{document}